\documentclass[12pt, preprint]{aastex}
\usepackage{emulateapj5}  

\begin{document}
\def\simeq{
\mathrel{\raise.3ex\hbox{$\sim$}\mkern-14mu\lower0.4ex\hbox{$-$}}
}


\def\lsun{{\rm L_{\odot}}}
\def\msun{{\rm M_{\odot}}}
\def\rsun{{\rm R_{\odot}}}
\def\lta{\la}
\def\gta{\ga}
\def\be{\begin{equation}}
\def\ee{\end{equation}}
\def\lsun{{\rm L_{\odot}}}
\def\le{{L_{\rm Edd}}}
\def\msun{{\rm M_{\odot}}}
\def\rsun{{\rm R_{\odot}}}
\def\rp{{R_{\rm ph}}}
\def\rs{{R_{\rm s}}}
\def\mo{{\dot M_{\rm out}}}
\def\me{{\dot M_{\rm Edd}}}
\def\tc{{t_{\rm C}}}
\def\rc{{R_{\rm core}}}
\def\mc{{M_{\rm core}}}
\def\mbh{{M_{\rm BH}}}
\def\e{{\dot m_{\rm E}}}

\title{HEATING CLUSTER GAS}

\author{ Andrew~King\altaffilmark{1}}

\altaffiltext{1} {Department of Physics and Astronomy, University of
Leicester, Leicester LE1 7RH, U.K.; ark@astro.le.ac.uk}

\begin{abstract}

It is now generally agreed that some process prevents the diffuse gas
in galaxy clusters from cooling significantly, although there is less
agreement about the nature of this process. I suggest that cluster gas
may be heated by a natural extension of the mechanism establishing the
$\mbh-\sigma$ and $\mbh - M_{\rm bulge}$ relations in galaxies, namely
outflows resulting from super--Eddington accretion on to the galaxy's
central black hole. The black holes in cD galaxies are sporadically
fed at unusually high Eddington ratios. These are triggered as the
cluster gas tries to cool, but rapidly quenched by the resulting shock
heating. This mechanism is close to the optimum efficiency for using
accretion energy to reheat cluster gas, and probably more effective
than `radio mode' heating by jets for example. The excess energy is
radiated in active phases of the cD galaxy nucleus, probably highly
anisotropically.

\end{abstract}


\section{Introduction}

Clusters of galaxies are the largest gravitationally bound objects in
the universe. Assuming rough virial equilibrium between the component
dark matter, gas and galaxies, with velocity dispersion $\sigma_c \sim
1000~{\rm km\,s^{-1}}$, shows that within a core radius $\rc \sim
150$~kpc about the central cD galaxy the intercluster gas (total mass
$\sim 10^{14}\msun$) has a free--free cooling timescale shorter than
the age of the Universe (see eq \ref{core} below). However it is by
now well established that there is no significant mass of cooling gas
within $\rc$ flowing towards the cD galaxy, implying that some
mechanism supplies energy to heat this gas. The most likely source of
this energy is fairly clear: the cD galaxy is very massive ($M_{\rm
  cD} \ga 10^{12}\msun$) and thus probably has a central black hole of
high mass $\mbh \ga 10^9\msun$. The total luminous accretion energy
$\epsilon Mc^2$ released in building up this hole mass (here $\epsilon
\sim 0.1$ is the radiative efficiency) considerably exceeds that
needed to resupply the energy lost in radiation by the gas within
$\rc$ (see Section 2 below).

However the means of transporting a suitable fraction of this energy
to the radiating gas is far less clear. Several methods have been
proposed, including sound waves and thermal conduction (e.g. Graham et al.,
2008; Conroy \& Ostriker, 2008) and mechanical
heating by jets (e.g. Br\"uggen \& Kaiser, 2002; Omma et al.,
2004). However it is uncertain how effective these mechanisms are in
coupling to the gas, particularly in view of the fact that the cD
galaxy is not active for most of the time.

A possible alternative heating mechanism comes from the relation
between central black hole mass $\mbh$ and {\it galaxy} velocity
dispersion $\sigma_g$ observed in nearby galaxies (Ferrarese \&
Merritt, 2000; Gebhardt et al., 2000; Tremaine et al., 2002), which
has the form $\mbh \propto \sigma_g^4$. Although alternative
explanations exist, a promising line (e.g. Silk \& Rees, 1998; King,
2003, 2005) suggests that this relation is a consequence of
super--Eddington accretion on to the central black hole. This drives
an outflow which communicates the hole's presence to the interstellar
gas. At modest Eddington factors $\e = \mo/\me$ the radiation field
couples to the outflow via the single--scattering limit. This imparts
momentum $\le/c$ to it (King \& Pounds, 2003; King, 2003, 2005), and
simultaneously Compton--cools the reverse shock as the outflow sweeps
up the galaxy's interstellar gas.  Thus a forward shock is driven into
the ambient gas purely by the momentum of the super--Eddington outflow
(`momentum--driven) with no extra contribution from the kinetic energy
randomised in the reverse shock (this would be an `energy driven'
flow).  The dynamics of this two--shock structure now fix the relation
between the black hole mass and the galaxy properties. If the black
hole mass is below a critical value $C\sigma_g^4$ the Eddington thrust
$\le/c$ is too weak to lift the interstellar gas against the galactic
potential measured by $\sigma_g$. Thus the shock does not propagate
outside the hole's immediate vicinity and accretion can
continue. However once $\mbh$ reaches the critical value
$C\sigma_g^4$, the shock attains the escape velocity $\sim \sigma_g$
and expands to large radii, preventing further growth in $\mbh$.

Remarkably, this simple idea gives not only the observed
proportionality $\mbh = C\sigma_g^4$, but also the quantitatively
correct coefficient $C = f_g\kappa/\pi G^2$, where $f_g \simeq 0.16$
is the gas fraction, $\kappa$ the electron--scattering cross--section,
and $G$ the gravitational constant (King, 2003, 2005). Further, at
sufficiently large distances $R_c$ from the hole, Compton cooling is
ineffective and the extra injection of thermalized kinetic energy
accelerates the two shocks above the escape velocity $\sigma_g$,
driving away the remaining gas and fixing the baryonic mass $M_{\rm
  bulge}$ of the galaxy bulge. In the limit of modest Eddington
factors (which implies wind outflow speeds $v$ approaching $c$) one
finds the relation $M_{\rm bulge} \sim \mbh (m_p/m_e)^2\sigma_g/c \sim
10^3\mbh$ between bulge and black hole mass, very close to
observation.

These results show that the outflows driven by super--Eddington
accretion are very effective in communicating the hole's presence to
the gas in the galaxy. Moreover if the outer (snowplow) shock reaches
a large distance $R_c$ from the hole it strongly heats this gas
because the shock velocity exceeds the local velocity dispersion. Here
we see a possible connection with the cluster gas cooling
problem. These features of super--Eddington outflows in galaxies are
obviously also desirable ingredients for any mechanism which might
heat {\it cluster} gas. If the accretion energy of the central black
hole could somehow drive a shell into the cluster gas, it could also
reach a radius where Compton cooling of the reverse shock is
ineffective.  At this point the shock velocities would increase
because some of the outflow kinetic energy is converted to heat and
hence exerts pressure. This higher shock speed would exceed the local
velocity dispersion in the cluster gas, heating the gas above the
virial temperature and thus supplying heat as well as kinetic energy
to the cluster gas.

However at first sight there appears to be a major difficulty in
extending this mechanism in this way. For at the expected mass $\mbh =
C\sigma_g^4$ the Eddington thrust of the hole is too weak to lift
cluster gas in the cluster potential measured by $\sigma_c > \sigma_g$
out to the radius $R_c$ where shock heating can be effective. This
would instead require the considerably larger black hole mass
$C\sigma_c^4 = (\sigma_c/\sigma_g)^4\mbh \sim 6\times
10^{10}\sigma_{1000}^4\msun$, where $\sigma_{1000} =
\sigma_c/1000~{\rm km\,s}^{-1}$.

But driving simply by the Eddington thrust $\le/c$ is a feature of
mildly super--Eddington accretion only. At higher Eddington ratios
$\e$, multiple scattering raises the thrust above this value, allowing
a hole with only the standard $\mbh-\sigma_g$ mass $C\sigma_g^4$ to drive
shocks into the stronger cluster potential and thus heat the cluster
gas out to the core radius $\rc$ (see Section 3).  This process
evidently has high efficiency in communicating accretion energy
released near the black hole to the distant cluster gas. We thus have
a potential explanation of the cluster heating problem as a natural
extension of the $\mbh -\sigma$ problem, provided that we can argue that
the black hole in the central cD galaxy should have an Eddington ratio
significantly larger than unity. Given its privileged position this
seems inherently plausible, and I discuss this in Section 4.

\section{Cluster Gas} 

To fix ideas, I derive here the properties of the core gas in a
typical cluster. For simplicity I approximate this as an isothermal
sphere characterised by the velocity dispersion $\sigma_c$. Then the
gas density at radius $r$ is
\begin{equation}
\rho = {f_g\sigma_c^2\over 2\pi Gr^2}
\label{rho}
\end{equation}
and the gas mass inside radius $R$ is
\begin{equation}
M(R) = 4\pi\int_0^R\rho r^2 {\rm d}r = {2f_g\sigma_c^2R\over G} \simeq
8\times 10^{13}\msun\sigma_{1000}^2R_{\rm Mpc}
\label{m}
\end{equation}
where $R_{\rm Mpc} = R/1~{\rm Mpc}$. Assuming virial equilibrium, the gas
temperature is $T \simeq 10^8\sigma_{1000}^2$~K. The dominant cooling process
is free--free emission, with cooling time $\propto T^{1/2}/\rho$. This is
shorter than a Hubble time $t_H$ for $\rho < \rho_{\rm cool} \simeq
10^{-26}\sigma_{1000}~{\rm g\, cm^{-3}}$, i.e. within a core radius (using eqn
\ref{rho})
\begin{equation}
\rc = 150\sigma_{1000}^{1/2}~{\rm kpc}
\label{core}
\end{equation}
From (\ref{m}) the mass of this core gas is
\begin{equation}
M_{\rm core} = 1.3\times 10^{13}\sigma_{1000}^{5/2}\msun
\label{mcore}
\end{equation}

To prevent significant cooling of this gas requires an energy input of
about 1~keV per baryon, i.e. an energy $E_h \sim (1~({\rm
  keV/10^3~MeV})M_{\rm core}c^2\sim 10^{-6}M_{\rm core}c^2$. The total
gravitational binding energy released in accreting mass $M_{\rm acc}$
on to the central black hole of the cD galaxy is $E_{\rm acc} = \epsilon M_{\rm
  acc} c^2$, with $\epsilon \sim 0.1$. If $\eta_{\rm heat}$ denotes the
efficiency with which this energy is used to heat the
cluster gas we see that the total accreted mass required to prevent
cooling is
\begin{equation}
M_{\rm acc,\, h} \simeq 10^{-5}{\mc\over (\epsilon/0.1)\eta_{\rm
    heat}} \simeq {1.3\times 10^8\sigma_{1000}^{5/2}\over
  (\epsilon/0.1)\eta_{\rm heat}}\msun.
\label{frac}
\end{equation}

Clearly heating by the central black hole cannot work if the required
mass $M_{\rm acc, \, h}$ exceeds its total mass $\mbh$. If this is $\sim
10^9\msun$ we need $\eta_{\rm heat} \ga 0.1$. The mechanism
described below has $\eta_{\rm heat} \simeq 0.2$. I compare this with
other forms of heating in Section 5.

\section{Heating Cluster Gas}

I suggest here that cluster gas may be heated by an extension of the
process establishing the $\mbh - \sigma_g$ and $\mbh - M_{\rm bulge}$
relations in individual galaxies, involving super--Eddington accretion
on to the central black hole.  The resulting outflow is roughly
spherical (see below) and sweeps up the galaxy gas into a thin shell
and tries to drive it out against gravity. Sijacki et al. (2007) have
recently performed a cosmological simulation with a form of mechanical
feedback on cluster gas, and indeed found that it could prevent
cooling. However much of the interaction between the outflow and the
cluster gas necessarily occurs on scales not accessible to current
numerical simulations. Here I adopt a simple analytic picture in the
hope of getting some physical insight into the process.

This type of approach is described in detail in King (2003, 2005). The
second of these papers gives the equation of motion of the swept--up
gas shell and shows that this clears the accreting gas away from the
central black hole once the hole mass reaches the critical value
\begin{equation}
M_{\sigma} = {f_g\kappa\over \pi G^2}\sigma_g^4.
\label{msig}
\end{equation}

One could follow the same procedure in considering the effects on
cluster gas, but for our purposes a simpler method is adequate. We
note that the weight of the shell of swept--up cluster gas is $W(R) =
GM(R)[M_{\rm total}(R)]/ R^2$: here $M(R)$ is the mass of the shell,
and $M_{\rm total}(R)$ is the total mass inside cluster radius $R$,
including dark matter. Neglecting the contribution of the cD galaxy
mass, which is small for $R \sim \rc$, this is simply $M(R)/f_g$,
since $M(R)$ is just the gas mass originally inside $R$ before the
passage of the shock. (Note that in King, 2005, the second term of eqn
(2) should read $GM(R)[M + M_{\rm total}(R)]/R^2$, and the correct
definition of $M_{\sigma}$ immediately below eqn (3) is
$(f_g\kappa/\pi G^2)\sigma^4$.)  Using (\ref{m}) we see that the
weight $W(R)$ is independent of $R$ for large $R$, and has the
constant value
\begin{equation}
W = {4f_g\sigma_c^4\over G}
\end{equation}
(the shell and total mass each increase as $R$, so their product exactly
cancels the inverse--square weakening of gravity). We can now decide whether
the shell reaches large $R$, and so heat the cluster gas, by comparing the
weight $W$ with the thrust produced by the super--Eddington accreting black
hole in the center of the cD galaxy (this procedure does not give the time
dependence of the motion, which requires one to solve the shell's equation of
motion taking account of its increasing inertia, cf King, 2005).

In the single--scattering limit expected for modest Eddington ratios
$\e$ this thrust is simply $\le/c = 4\pi GM/\kappa$. This gives the
expected result that the shell would reach large $R$, and thus heat
the cluster gas, if and only if the black hole mass exceeded the value
(\ref{msig}) with $\sigma_c$ in place of $\sigma_g$, which as we have
seen in Section 1 is unrealistically large ($\sim 6\times
10^{10}\msun$). We would expect instead that as in other galaxies, the
hole would have only reached the smaller value (\ref{msig}) given by
the galaxy's internal velocity dispersion $\sigma_g$.

Now let us consider the effects of an accretion episode with higher
Eddington ratio. Shakura \& Sunyaev (1973) consider disc accretion in
this case. Their theory appears to apply well to X--ray binary
systems, which can have far higher Eddington ratios than supermassive
black holes in galaxy centers (see eq \ref{e} below). For the
well--known system SS433, which has $\e \sim 5000$, Begelman et al.
(2006) and Poutanen et al. (2007) show that the features anticipated
by Shakura \& Sunyaev appear. In particular the total accretion
luminosity is $\simeq\le[1 + \ln\e]$, and is almost entirely
channelled by scattering into a narrow pair of funnels around the disc
axis, so that the outflow is essentially spherical apart from these
two funnels. These two results suggest that highly super--Eddington
accretion on to stellar--mass compact objects offers a plausible
explanation for most if not all ultraluminous X--ray sources (ULXs: cf
King et al., 2001; King, 2009).  Most importantly for our purposes,
most of the super--Eddington mass inflow is blown away from a radius
$R_{\rm circ} \simeq 9\e R_{\rm in}/4$ (where $R_{\rm in}$ is the
inner disc radius near the black hole) with mechanical luminosity
\begin{equation}
{1\over 2}\dot Mv^2 \simeq \le
\end{equation}
This resulting relation $v = (2\le/\dot M)^{1/2}$ allows us to estimate
the thrust exerted by the accreting hole on its
surroundings, i.e.
\begin{equation}
\dot Mv = (2\dot M\le)^{1/2} = \left({2\dot M\over \le}\right)^{1/2}\le = 
\left({2\e\over \epsilon}\right)^{1/2}{\le\over c}
\label{thrust}
\end{equation}
This exceeds the single--scattering estimate by the factor $\sim
(2\e/\epsilon)^{1/2}$. This is potentially a lower limit to the
increase, as there is a thermal pressure contribution if the external
shock cannot cool. However Compton cooling still operates on the
outflow, since a luminosity $\sim \le$ escapes isotropically through
the outflow rather than via the funnels along the disc axis. This also
means that the temperature profile remains flat or decreasing radially
inwards in the cluster center.

The estimate (\ref{thrust}) shows that the likely black hole mass
results in a large enough thrust to heat the cluster gas if a
significant Eddington ratio holds for some fraction of a cluster
dynamical time $R_{\rm core}/\sigma_c \sim 1.5\times
10^8$~yr. Equating the value (\ref{thrust}) to the weight
$4f_g\sigma_c^4/G$ of the cluster gas shows that the minimum black
hole mass needed to get the shock out to the core radius is
\begin{equation}
\mbh = \left({\epsilon\over 2\e}\right)^{1/2}{f_g\kappa\over \pi G^2}\sigma_c^4.
\end{equation}
Put another way, a super--Eddington accretion episode can successfully heat
the cluster gas provided that the Eddington ratio exceeds the critical value
\begin{equation}
\e({\rm crit}) = {180\epsilon_{0.1}\over M_9^2}\sigma_{1000}^4
\label{crit}
\end{equation}
for a fraction of the cluster dynamical time, where $\epsilon_{0.1} =
\epsilon/0.1$ and $M_9 = \mbh/10^9\msun$.

\section{Accretion}

We have seen that a sufficiently high Eddington ratio for the central
black hole in the cD galaxy is required to heat cluster gas, albeit
for a relatively short timescale. The accretion rate specified by
(\ref{crit}) is extreme -- close to the dynamical rate $\sim
f_g\sigma_g^3/G$ for the host cD galaxy. If the black hole in the
latter obeys the $\mbh-\sigma_g$ relation, this rate implies an
Eddington ratio
\begin{equation}
\e \simeq {\epsilon c\over 4\sigma_g} \sim 30
\label{e}
\end{equation}
where I have taken $\epsilon = 0.1, \sigma_g = 300~{\rm km\,s^{-1}}$ at the last
step. This is in one sense reassuring, as it shows that most supermassive black
holes in galaxy centers do not experience very high Eddington
ratios, justifying the use of the single--scattering limit in deriving the
$\mbh -\sigma$ relation for them. Conversely, if cluster gas is heated by the
process discussed here, the central regions of the cD galaxy must experience
gas inflow rates of order $10^3\msun\, {\rm yr}^{-1}$, which come close to
destabilizing them, at least for a short time.

There is an obvious candidate for this very rapid accretion -- the
onset of the cooling catastrophe. If nothing acted to reheat the
cluster gas, this would eventually begin to flow in towards the cD
galaxy at rates $\sim \mc/t_H \sim 10^3\sigma_{1000}^{5/2}~\msun\,
{\rm yr}^{-1}$. The cD galaxy must react long before such rates are
reached. Its central black hole drives a snowplow shock out through
the cluster gas, reaching the core radius and reheating the enclosed
gas to the virial temperature in a dynamical time $\sim 10^8$~yr. This
stabilizes the cluster gas and stops further infall. The duty cycle of
the cooling--infall phase is thus of order $10^{-2}$. Only a small
amount of cluster can cool before being reheated, so clusters
typically appear to be in virial equilibrium.

\section{Energy Budget}

During an active phase of the type described above, the central black
hole gains mass at about its Eddington rate for some $10^8$~yr, and
thus typically grows by $\sim 10^9~\msun$. In return it reheats $\sim
10^{13}\msun$ of cluster gas. Comparing with the estimate
(\ref{frac}), this process uses about 5 -- 10 times more than the
minimum possible accretion fraction of $10^{-5}$. The reason for this
is that the central black hole manages to radiate about 4 times the
energy it puts into mechanical luminosity (respectively $\le[1 +
\ln\e] \sim 4\le$, and $\le$) and more energy is lost in cooling. The
cD galaxy gains $\sim 10^{11}\msun$ during an active phase, small
compared with its current mass. We can thus write
\begin{equation}
E_q \simeq 0.1\epsilon c^2M_q
\end{equation}
for the heat input into the cluster in this form of the `quasar' mode
from accreting mass $M_q$ on to the black hole.  Discussions of
cluster heating (e.g. Sijacki et al., 2007) contrast the quasar mode
with the `radio' mode. This is motivated by observations (e.g. Birzan
et al., 2004, Rafferty et al., 2006) which suggest that radio--loud
FR~I sources can inflate X--ray cavities. The accretion of gas
mass $M_r$ is assumed to produce jet kinetic energy
\begin{equation}
E_j = \eta_r c^2M_r.
\end{equation}
If these jets convert their energy into cluster
heating with efficiency $\eta_j$ we get heat input
\begin{equation}
\eta_r\eta_j c^2 M_r
\end{equation}
into the cluster gas. Hence
\begin{equation}
{E_r\over E_q} \simeq {\eta_r\eta_jM_r\over 0.1\epsilon M_q}.
\label{heat}
\end{equation}
However several observational surveys put strict limits on the ratio
of total jet to radiative output by AGN, or equivalently
$\eta_rM_r/\epsilon M_q$. For example Cattaneo \& Best (2009; see also
Merloni \& Heinz, 2008)
find this ratio is $\la 0.1$. Accordingly we find from
(\ref{heat}) that
\begin{equation}
{E_r\over E_q} \la \eta_j < 1.
\label{j}
\end{equation}
This suggests that outflows of the type considered here likely to be
more effective than jets in heating cluster gas, i.e. require less
black hole mass growth to produce the same heating effect.  Thus for
heating by an outflow the total increase $\Delta\mbh$ in the mass of
the central black hole is controlled by the rate at which cluster gas
cools, i.e. $\Delta\mbh \simeq 10^{-4}M_{\rm cool}$, so if $M_{\rm
  cool} \sim M_{\rm core}$ we expect $\Delta\mbh \sim 10^9\msun$. In
principle $M_{\rm cool}$ might exceed $M_{\rm core}$ if the reheated
gas cools more quickly than before, i.e. in less than a Hubble
time. This would then require multiple heating events, and thus black
hole mass growth $\mbh > 10^9\msun$. In the simple
spherically--symmetric picture adopted here this does not happen, but
this conclusion should be checked by numerical simulations allowing
for deviations from this symmetry and thus local cooling
instabilities.  In view of eqn (\ref{j}), if this picture requires
excessive black hole mass growth, this problem is likely to be worse
for radio mode heating.

\section{Discussion}

I have suggested that cluster gas is heated by a natural extension of the
process establishing the $\mbh -\sigma$ and $\mbh - M_{\rm bulge}$ relations in
galaxies. The privileged position of the central cD galaxy means that it is
intermittently subject to extremely high gas inflow rates. These trigger
highly super--Eddington accretion on to the central black hole, which reacts
by driving a shock into the infalling gas, efficiently reheating it and
stabilizing the cluster gas. The duty cycle of an active phase of this type is
about $10^{-2}$, so most clusters appear to be stably in virial
equilibrium. 

During an active phase the accreting central black hole of the cD
galaxy emits $\sim 4\le \sim 4\times 10^{47}~{\rm erg\, s^{-1}}$ into
a narrow pair of cones, and $\sim \le \sim 10^{47}~{\rm erg\, s^{-1}}$
isotropically. If the scaling of the beaming factor with Eddington
ratio derived by King (2008) for ULXs holds here too, an observer
situated in these cones would infer a still higher apparent luminosity
$\sim 0.1\e^2\times 4\le \sim 4\times 10^{48}~{\rm erg\,
  s^{-1}}$. However as the cone solid angle is only $\sim 10^{-2}$,
and the duty cycle of active phases is also $10^{-2}$, it is
unsurprising that such luminosities are not observed. An active phase
in a cD galaxy would be observable through the isotropic Eddington
emission. The dense outflowing wind implies a large photosphere,
shifting the emission into the infrared, and a systematic search here
might prove interesting.

Comparison of the simple treatment given here with observation
requires care. In particular a more realistic cluster potential must
affect the ability of the swept--up gas shell to escape, and thus the
duration of the active phases and the temperature structure of the
cluster gas. Local density perturbations and the resulting cooling
will have similar effects. These could produce shocked bubbles whose
cooling times are significantly shorter than a Hubble time, as appears
to be true of at least some observed cases. One would then require
multiple heating events, totalling a much larger fraction of the
cluster lifetime, in order to stave off catastropic cooling, rather
than the $\sim 1$\% total heating time envisaged here. Answering these
questions requires numerical simulation.

\acknowledgements

I thank Sergei Nayakshin, Chris Power, Debora Sijacki, Walter Dehnen,
Mark Wilkinson, and Gordon Stewart for illuminating discussions, and
the referee for a very helpful report. Research in theoretical
astrophysics at Leicester is supported by an STFC Rolling Grant.

\end{document}